\documentstyle[preprint,aps,eqsecnum]{revtex}

\begin{document}
\draft
\title{Analytic approach to the polarization of the\\
cosmic microwave background in flat and open universes}
\author{Mat\'\i as Zaldarriaga and Diego D. Harari}
\address{Departamento de F{\'\i}sica, Facultad de Ciencias
Exactas y Naturales\\ Universidad de Buenos Aires \\ Ciudad Universitaria -
Pab. 1\\ 1428 Buenos Aires, Argentina}
\maketitle
\begin{abstract}
\tighten
We develop an analytic method and approximations to compute the polarization
induced in the cosmic microwave background radiation on a wide range of
angular scales by anisotropic Thomson scattering in presence of adiabatic
scalar {(energy-density)} linear fluctuations. The formalism is an extension
to the polarized case of the analytic approach recently developed by Hu and
Sugiyama to evaluate the (unpolarized) temperature correlation function. The
analytic approach helps to highlight the dependence of potentially measurable
polarization properties of the cosmic microwave background radiation upon
various parameters of the cosmological model. We show, for instance, that the
ratio between the multipoles of the temperature and polarization correlation
functions depends very sensitively upon the value of $\Omega_0$, the matter
density in units of the critical, in an open universe.
\end{abstract}
\pacs{PACS numbers: 98.80.-k, 98.70.Vc, 98.80.Es}

\section{Introduction}

The cosmic microwave background radiation (CMB) is a seemingly unlimited source
of information about the history and evolution of the Universe. Its existence
is one of the most powerful arguments for the standard ``hot'' big-bang
cosmology. Its high degree of isotropy is an indication of the Universe
large scale homogeneity. Its precise black-body spectrum places stringent
bounds on alternative cosmologies. Its large scale anisotropies uncover
tiny fluctuations in the gravitational potential, most likely
the same that once upon a time led to the formation of galaxies and other
large scale structure in the Universe.

The polarization properties of the cosmic microwave background radiation
constitute yet another set of observables whose eventual measurement is still
to bear fruit. A positive measurement of a degree of polarization
would provide much further insight into the Universe history and evolution.
The current bound on  the degree of linear polarization of the CMB  on large
angular scales, $P<6\times 10^{-5}$ \cite{Lubin83}, already serves to set
limits and constraints upon alternative cosmological scenarios.

The fact that anisotropic Thomson scattering of photons and electrons around
the time of decoupling induces a degree of linear polarization in the cosmic
microwave background\cite{Rees68} was pointed out shortly after its discovery
in 1965. Several estimates were made of the degree of polarization expected.
The task is to evaluate the Stokes parameters of the photon distribution
function, which satisfies a Boltzmann equation with a Thomson scattering
collisional term\cite{Chandrasekar50}. The first computations were performed
in homogeneous but anisotropically expanding
universes\cite{Rees68,Nanos79,Basko79,Negroponte80,Tolman84,Tolman85}.
{Energy-density} fluctuations
\cite{Bond84,Milaneschi86,Bond87,Crittenden93,Harari93,Coulson94,Ng95} and
long wavelength gravitational waves
\cite{Polnarev85,Frewin93,Crittenden93,Harari93,Ng94} were also considered as
the source of the anisotropy that leads to polarization.  Typically, at least
within standard recombination histories, the degree of linear polarization
turns out to be almost two orders of magnitude smaller than the large scale
temperature anisotropy. This is still well beyond current detection
capabilities. Nevertheless, with large scale anisotropies in the CMB now
positively measured by COBE-DMR\cite{COBE92} and other experiments, it is
worthwhile to further analyse the predictions for the CMB polarization
properties within alternative realistic cosmological models, and its
dependence upon various parameters.

Very detailed numerical computations of the CMB temperature fluctuations and
polarization correlation functions on all angular scales were performed by
Bond and Efstathiou\cite{Bond87} for standard $\Omega_0=1$ cold dark matter
dominated universes, with scale invariant adiabatic or isocurvature scalar
fluctuations. It is one of the purposes of the present paper to reproduce the
most significant features of the multipole expansion of the polarization
correlation function on all angular scales in this cosmological model from an
analytic approach, to better highlight its dependence upon various parameters.
To achieve that goal, in Section II we develop a method and perform
approximations that very closely parallel the recent analytic approach
introduced by Hu and Sugiyama to evaluate the temperature anisotropies on all
angular scales for the unpolarized case\cite{Hu94}. In Section III the
multipole expansion of the polarization correlation function is performed,
and an expression for the total degree of linear polarization is derived. In
Section IV we specialize to a specific cosmological model, namely a spatially-
flat, $\Omega_0=1$ cold dark matter cosmology with scale invariant scalar
fluctuations normalized to the large angle anisotropy measured by COBE, under
the assumption that the baryonic energy density is much smaller than the
radiation energy density at the time of decoupling. In Section V we find the
functional dependence of the ratio between the multipoles of the temperature
fluctuation and of the polarization correlation functions upon  $\Omega_0$,
the mass-density in units of the critical, both in open universes as well as
in spatially-flat models with a non-vanishing cosmological constant, with
scale-invariant scalar fluctuations as the source of the anisotropy. Section
VI rounds up the conclusions.

\section{Analytic approach}
\subsection{Tight Coupling approximation}

The Boltzmann equations for the photon distribution function in realistic cold
dark matter cosmological models with energy-density linear fluctuations have
been numerically solved, and the predictions for the CMB temperature
correlation function on all angular scales have been analised in great
detail\cite{Bond84,Vittorio84,Bond87}. The Boltzmann code was also used to
study the dependence of the microwave background anisotropies upon various
cosmological parameters\cite{Bond94}. Notwithstanding the thoroughness of
these numerical studies, it is very useful to have some analytic approximation
to the exact solution, to gain even further insight into the nature and origin
of the CMB anisotropies. Some analytic methods and approximations to compute
with reasonable accuracy the CMB anisotropies on all angular scales were
recently developed\cite{Hu94,Seljak94,Jorgensen94}, improving upon the
original work of Sachs and Wolfe\cite{Sachs67}.

Here we extend to the polarized case the analytic approach that Hu and
Sugiyama introduced to compute CMB anisotropies\cite{Hu94}. The method is in
turn an extension of the standard tight coupling
analysis\cite{Peebles70,Doroshkevich78} to include realistic time dependent
gravitational potentials, described in a  gauge invariant
formalism\cite{Bardeen80,Kodama84,Mukhanov92}.
It is based upon an expansion of the CMB
temperature fluctuation in inverse powers of the differential optical depth.
In the tight coupling regime, when the effectiveness of Thomson scattering of
CMB photons with free electrons makes the differential optical depth high, a
perturbative expansion to first order constitutes a very good approximation to
the exact result. While photons and baryons are tightly coupled, all higher
multipoles of the temperature fluctuation can be evaluated in terms of the
monopole, which in turn obeys the equation of a forced and damped
oscillator\cite{Hu94}. We will see now how to include the polarization
dependence of Thomson scattering into this formalism.

The CMB gauge-invariant temperature fluctuation for a given direction  of
observation, described in terms of polar angles $\theta,\phi$, is  the
relative temperature fluctuation around the mean evaluated in the shear-free
(Newtonian) gauge: $\Delta_T(\theta,\phi)\equiv\Delta T(\theta,\phi)/T$. We
compute first the temperature fluctuations induced by only one Fourier mode of
the scalar fluctuations in the gravitational potential, with wave-vector $\vec
k$. Then it is convenient to use a reference frame such that $\hat
z\parallel\vec k$, since there is axial symmetry around the direction of $\vec
k$. The degree of linear polarization, $\Delta_P$, is defined in terms of the
Stokes parameters $Q$ and $U$\cite{Chandrasekar50} of the CMB radiation. We
choose the two orthogonal directions into which the intensity is projected to
define the Stokes parameters as $\hat\theta$ and $\hat\phi$ of spherical
coordinates. The advantage is that, in this basis, $U=0$, and the only non-
vanishing Stokes parameter is $Q= I_{\hat\theta} - I_{\hat\phi}/I_{\hat\theta}
+ I_{\hat\phi}\equiv\Delta_P$\cite{footnote1}.

Given the axial symmetry around the direction of $\vec k\parallel\hat z$, the
multipole expansion of either the temperature fluctuation or the polarization
in a given direction of observation $\hat n$ can be written as
\begin{equation}
\Delta(\hat n,\vec k)=\sum_l (2 l+1) \Delta_l P_l(\mu)
\end{equation}
with $P_l$ the Legendre polynomials and $\mu\equiv
\cos\theta=\vec k\cdot\hat n/k$\cite{footnote2}.

In a spatially-flat Robertson-Walker metric with linear density fluctuations
described by gauge invariant potentials $\Phi,\Psi$, and after angular
integration of the collisional term of the Boltzmann
equations\cite{Chandrasekar50}, made easier by the axial symmetry, the
evolution  equations for the Fourier mode of wavevector $\vec k$ of the
gauge-invariant temperature fluctuation and polarization
read\cite{Hu94,Bond87,Kodama86}
\begin{equation} \begin{array}{lll}
&\dot\Delta_T +ik\mu (\Delta_T+\Psi) =-\dot\Phi-\dot\kappa\{\Delta_T -
\Delta_{T0} -\mu V_b -{1\over 2}P_2(\mu)[\Delta_{T2}+\Delta_{P2}-
\Delta_{P0}]\}\\ & \\ &\dot\Delta_P +ik\mu \Delta_P = -\dot\kappa \{\Delta_P +
{1\over 2} [1-P_2(\mu)] [\Delta_{T2}+\Delta_{P2}-\Delta_{P0}]\}\\ \end{array}
\label{Boltzmann}
\end{equation}
where a dot means derivative with respect to the conformal time $\tau=\int dt
a_o /a$, with $a(t)$ the scale factor of the spatially-flat Robertson-Walker
metric, and $a_o=a(t_o)$ its value at present time. We shall write in this
work the present value of the Hubble coefficient as $H_0=h100{\rm km.s}^{-1}.
{\rm Mpc}^{-1}$.
$\dot\kappa=x_e n_e \sigma_T a/a_0$ is the differential optical depth for
Thomson scattering, with $x_e$ the fraction of ionized electrons with number
density $n_e$, and $\sigma_T$ the Thomson scattering cross section.  $V_b$ is
the velocity of the baryons. $\Phi$, $\Psi$ are the Fourier modes of the
gravitational potentials\cite{Bardeen80,Mukhanov92}.

The equation of motion for the baryons reads
\begin{equation}
\dot V_b= -{\dot a\over a}V_b -i k \Psi+{\dot\kappa\over R}(3\Delta_{T1}-V_b)
\end{equation}
where $R\equiv 3\rho_b /4\rho_{\gamma}$, the ratio between baryonic
and radiation densitites.

Equations (\ref{Boltzmann}) can be formally integrated
\begin{equation}\begin{array}{rl}
(\Delta_T +\Psi)&=\int_0^{\tau_0}d\tau e^{i k \mu (\tau -\tau_0)}
e^{-\kappa(\tau_0,\tau)}  \\
&\quad\quad\quad
\{ \dot\kappa (\Delta_{T0}+\Psi+\mu V_b + {1\over 2} P_2(\mu)
[\Delta_{T2}+\Delta_{P2}-\Delta_{P0}])-\dot\Phi+\dot\Psi\}\\
& \\
\Delta_P &= -\int_0^{\tau_0} d\tau e^{i k \mu (\tau -\tau_0)}
\dot\kappa e^{-\kappa(\tau_0,\tau)} {1\over 2} [1-P_2(\mu)]
[\Delta_{T2}+\Delta_{P2}-\Delta_{P0}]\quad ,\\
\end{array}\label{formal}
\end{equation}
where
\begin{equation}
\kappa(\tau_0,\tau)=\int_\tau^{\tau_0}x_en_e\sigma_T{a(\tau)\over
a(\tau_0)}d\tau
\label{opticaldepth}
\end{equation}
is the optical depth to photons emitted at conformal time $\tau$. The
combination $\dot\kappa e^{-\kappa}$ is called the conformal time visibility
function. It is the probability that photons last scattered within $d\tau$ of
$\tau$. For standard recombination this function has a sharp peak at the
conformal time of decoupling $\tau_D$\cite{Bond88}. Thus, the integral for
$\Delta_P$ in eq. (\ref{formal}) is dominated by the value of the integrand
around decoupling. In other words, for standard recombination histories, with
no reionization, the polarization of the CMB we observe today was produced
just before decoupling.

In order to approximately solve the integral for $\Delta_P$ in eq.
(\ref{formal}), we need to know the value of the combination $S_P\equiv
[\Delta_{P0}-\Delta_{T2}-\Delta_{P2}]$ around decoupling. We can evaluate the
first multipoles of the CMB temperature fluctuations and polarization before
decoupling in the tight coupling approximation\cite{Hu94}, {\it i.e.} by a
perturbative expansion in inverse powers of the differential optical depth
$\dot\kappa$, which is high before decoupling. In order to perform this
perturbative expansion we first rewrite the evolution equations
(\ref{Boltzmann}) in the following form
\begin{equation}\begin{array}{ll}
&\Delta_T -\Delta_{T0}-\mu V_b -{1\over 2}P_2(\mu)(\Delta_{T2}+\Delta_{P2}
-\Delta_{P0})= -\tau_C [\dot\Delta_T +ik\mu (\Delta_T+\Psi)+\dot\Phi ]\\
& \\
& \Delta_P +{1\over 2} (1-P_2(\mu))[\Delta_{T2}+\Delta_{P2}
-\Delta_{P0}]=-\tau_C (\dot\Delta_P +ik\mu \Delta_P)\\
& \\
&3 \Delta_{T1}- V_b=\tau_C R [a^{-1} {d \over d\tau}(a V_b)+i k \Psi]\\
\end{array}
\label{BoltzmannC}
\end{equation}
and expand all quantities of interest in powers of $\tau_C \equiv
\dot\kappa^{-1}$, the conformal time between Compton (or Thomson) scatterings,
assumed large for the tight coupling to be a sensible approximation.

In the strict tight coupling limit, that is to order zero in $\tau_C=\kappa^{-
1}$, the solutions to these equations read
\begin{equation}\begin{array}{ll}
&\Delta_{T1}={1\over 3}V_b\quad ;\quad
\Delta_{Tl}=0 \quad {\rm if}\quad l\ge 2\\
&\\
&\Delta_P=0\quad .
\end{array}\end{equation}
The interpretation of these formulae is very simple. In the lowest order
approximation the photons and baryons are so strongly coupled that the photon
distribution is isotropic in the baryon's rest frame. The photon distribution
being isotropic, Thomson scattering does not polarize the CMB.

Now we expand eqs. (\ref{BoltzmannC}) to first order in
$\tau_C\equiv\dot\kappa^{-1}$, and get
\begin{equation}\begin{array}{ll}
&\Delta_{P2}=-{1\over 5}\Delta_{P0}={1\over 4}\Delta_{T2}\quad ;
\quad\Delta_{T2}=-{8\over 15}ik\tau_C\Delta_{T1}\\
&\\
&\Delta_{T1}={i\over k}(\dot\Delta_{T0}+\dot\Phi)\\
&\\
&\Delta_{Tl}=\Delta_{Pl}=0\quad {\rm if}\quad l\ge 3\quad ;
\quad\quad\Delta_{P1}=0\quad  .\\
\end{array}\label{firstorder}\end{equation}
These equations also have a simple interpretation. The polarization of the CMB
is proportional to the quadrupole of the photon distribution function (a dipole
does not induce polarization). The quadrupole in the temperature
fluctuation, in its turn, is produced by the ``free streaming'' of the dipole
between collisions. We see this from the relation
$\Delta_{T2}\propto k\tau_C\Delta_{T1}$. The tight coupling approximation is
actually valid when $k\tau_C<<1$, {\it i.e.} for wavelengths much larger than
the photon mean free path. The dipole in the temperature fluctuation can be
derived from the gravitational potential and the monopole, through  the
relation $i k \Delta_{T1} = -(\dot\Delta_{T0}+\dot\Phi)$. The monopole itself
obeys, in the tight coupling limit, the equation of a forced and damped
oscillator\cite{Hu94}
\begin{equation}
\ddot \Delta_{T0}+{\dot R\over
1+R}\dot\Delta_{T0}+ {k^2 \over 3(1+R)}\Delta_{T0}=-\ddot \Phi-{\dot R\over
1+R}\dot\Phi -{k^2 \Psi \over 3}
\label{monopol2}
\end{equation}
The solution to this equation, for a given cosmological model, explains the
properties of the CMB anisotropies\cite{Hu94}, and will also determine its
polarization properties, as we show in the next subsection. We postpone to
Section IV the solution of eq. (\ref{monopol2}) for a specific, realistic
cosmological model, and the analysis of its most significant features, such as
Doppler peaks, etc.

\subsection{Polarization}

Here we evaluate the integral for the polarization $\Delta_P$ in eq.
(\ref{formal}), which can be rewritten as:
\begin{equation}
\Delta_P ={3\over 4} (1-\mu^2)
 \int_0^{\tau_0} d\tau e^{i k \mu (\tau -\tau_0)}
\dot\kappa e^{-\kappa(\tau_0,\tau)}
S_P
\label{dp}
\end{equation}
where we have defined
\begin{equation}
S_P\equiv  \Delta_{P0}-\Delta_{T2}-\Delta_{P2}\quad .
\end{equation}
Because the visibility function ${\dot \kappa} e^{-\kappa(\tau_0,\tau)}$ is
strongly peaked around the time of decoupling, $\tau_D$, to calculate the
polarization today it is only necessary to know $S_P$ near decoupling. Right
before decoupling, and to first order in the tight coupling expansion, valid
for scales such that $k\tau_C<<1$, $S_P$ can be approximated, using eqs.
(\ref{firstorder}), as
\begin{equation}
S_P\approx {4\over 3} i k\tau_C \Delta_{T1} \quad.
\end{equation}
But since $\tau_C=\kappa^{-1}$ grows very fast  during recombination, we need
to know the time dependence of $S_P$ around decoupling with better
approximation than this. From (\ref{Boltzmann}) we find that $S_P$ satisfies
the following equation
\begin{equation}
\dot S_P+{3\over 10} \dot\kappa S_P=i k [{2\over 5}\Delta_{T1}
+{3\over 5}
(\Delta_{T3}+ \Delta_{P3}-\Delta_{P1})]
\end{equation}
During recombination we can approximate the right hand side of this equation
by its tight coupling expansion to first order, and thus keep just
$\Delta_{T1}$, as given by eq. (\ref{firstorder}). In this case the approximate
solution for $S_P$ is
\begin{equation}
S_P(\tau)={2\over 5} i k \int_0^{\tau} d\tau^{\prime}
 \Delta_{T1} e^{-{3\over 10} \kappa(\tau,\tau^{\prime})}
\end{equation}
We now replace this in the integrand of eq. (\ref{dp}) for $\Delta_P$. The
integral is dominated by the contribution around decoupling, since the
visibility function is strongly peaked around $\tau_D$, the conformal time of
decoupling. The conformal time visibility function is well approximated by a
gaussian, of width $\Delta\tau_D$\cite{Jones85,Bond88}. This means that
photons were able to travel a distance of order  $\Delta\tau_D $ between their
last two     scatterings. This is the time the quadrupole had to grow, and
thus the final polarization should be proportional to $k\Delta\tau_D
\Delta_{T1}$. To see that this is indeed the case, we perform the integrals
leading to $S_P(\tau)$ around decoupling and to $\Delta_P$ under the following
approximations, analogous to those in refs.\cite{Polnarev85,Harari93}. We
first approximate $\dot\kappa(\tau_0,\tau)\approx -{\kappa(\tau_0,\tau)\over
\Delta\tau_D}$, which is justified by the gaussian nature of the visibility
function  during recombination. Notice also that $\kappa(\tau,\tau^{\prime})=
\kappa(\tau_0,\tau^{\prime})-\kappa(\tau_0,\tau)$. We also neglect the time
variation of $\Delta_{T1}$ during the decoupling transition. Then we can
write, for $\tau$ around decoupling:
\begin{equation}
S_P(\tau)\approx{2\over 5}ik\Delta_{T1}(\tau_D)\Delta\tau_D e^{{3\over
10}\kappa(\tau_0,\tau)}\int_1^\infty {dx\over x}e^{-{3\over 10}x\kappa}\quad ,
\end{equation}
where we have changed the integration variable from $\tau'$ to
$x=\kappa(\tau_0,\tau)/\kappa(\tau_0,\tau^\prime)$. Now, neglecting also the
time variation of $e^{i k \mu (\tau -\tau_0)}$ during recombination, we get
\begin{equation}\begin{array}{lll}
&\Delta_P &={3\over 4} (1-\mu^2) e^{i k \mu (\tau_D -\tau_0)} {2\over 5}ik
\Delta_{T1}(\tau_D) \Delta\tau_D\int_0^{\infty} d\kappa e^{-{7\over 10}\kappa}
\int_1^\infty {dx\over x}e^{-{3\over 10}x\kappa} \\
& & \\
&  &=(1-\mu^2) e^{i k \mu (\tau_D -\tau_0)} 0.51 i k \Delta_{T1}(\tau_D)
\Delta\tau_D \equiv (1-\mu^2) e^{i k \mu (\tau_D -\tau_0)}\beta(k)
\quad . \\
\end{array}\label{soldp}
\end{equation}
We here defined, for shortness and later reference, the quantity $\beta (k)$.
Expression (\ref{soldp}) is one of our main analytic results. It gives,
for standard recombination histories, the polarization induced upon the CMB
by one Fourier mode of wavevector $\vec k$ of the linear density fluctuations,
in terms of the value of the dipole in the total temperature fluctuation at the
time of decoupling. It is proportional to the width of the last scattering
surface, $\Delta\tau_D$, because it is actually the quadrupole in the
temperature fluctuation during the last few scatterings what induces the
polarization, and the quadrupole itself is proportional to the dipole times
the width of the last scattering surface.

Expression (\ref{soldp}) is strictly valid only for scales such that
$k\Delta\tau_D<<1$, since we took the exponential out of the integral and
simply evaluated it at $\tau=\tau_D$. For scales such that $k\Delta\tau_D>>1$,
the oscillations in the integrand produce a cancellation. In other words, the
finite thickness of the last scattering surface damps the final polarization
on these scales. The degree of polarization is thus largest for modes of
wavelength comparable to  $\Delta\tau_D$.

In the long wavelength limit, the expression (\ref{soldp}) reduces to our
previous analytic estimate of the polarization on large angular
scales\cite{Harari93}, based on the method developed by
Polnarev\cite{Polnarev85}, once the dependence of $\Delta_{T1}$ with the
gravitational potentials is replaced.

The polarization properties of the CMB are very sensitive to the details of
the ionization history, and could very well serve to trace it
back\cite{Basko79,Naselskii87}. The proportionality in $\Delta\tau_D$ in our
solution is a hint of this. In a scenario with an appreciable reionization,
the polarization would increase, because the quadrupole of the temperature
anisotropy in the electron's rest frame, which is the source of polarization,
would be greatly enhanced.

A very important conclusion that can be drawn from eq. (\ref{soldp}) is that
one can determine the value of the dipole of the temperature fluctuations at
recombination measuring the present polarization properties of the CMB
radiation, at least if there was no reionization after recombination. This is
a very interesting perspective, since it could also serve, for instance, to
test alternative evolutions after recombination. The value of the dipole at
recombination, $\Delta_{T1}(\tau_D)$, can be derived from the monopole and the
gravitational potentials using eq. (\ref{firstorder}), and the monopole itself
solving eq. (\ref{monopol2}), in the tight coupling approximation. The
oscillatory behaviour of $\Delta_T$ corresponds to the so-called Doppler
peaks. In the case of adiabatic density fluctuations, the monopole turns out
to be proportional to $\cos\phi$ with $\phi=k\int^\tau_0 d\tau c_s$, where
$c_s=[3(1+R)]^{-1/2}$ is the photon-baryon fluid sound speed, while the dipole
is proportional to $\sin\phi$\cite{Hu94}. Thus, in the case of adiabatic
perturbations the peaks in the polarization $\Delta_P$ are located at
wavevectors such that $\phi(\tau_D)=(m+{1\over 2})\pi$ with $m$ an integer.
For models with low $\Omega_b$ where $c_s\sim {\rm
const.}\sim{1\over\sqrt{3}}$, the peaks are at ${k\tau_D\over\sqrt{3}}
=(m+{1\over 2}) \pi$. In the case of isocurvature perturbations the monopole
is proportional to $\sin\phi$, and the peaks in the polarization are instead
at $\phi(\tau_D)= m \pi$. A test that the relative locations of the peaks in
$\Delta_T$ and $\Delta_P$ verify these relations may serve as a test if the
recombination process was the standard one assumed here or not.

The relative heights of the different peaks is also a potential source of
information about cosmological parameters. As can be seen from eq.
(\ref{formal}), when $\Delta_{T0}$ and $\Psi$ have opposite sign a supression
in the height of the corresponding peak in $\Delta_T$ may occur. In the case
of adiabatic fluctuations, this can happen to even peaks. This pattern of
suppresion of even peaks relative to odd ones is sensitive to the value of
$\Omega_bh^2$: the suppression grows with $\Omega_bh^2$.\cite{Hu94} This
pattern of relative suppresion in the heights of even peaks does not occur to
the polarization.  This is so because the gravitational infall represented by
$\Psi$, which affects the anisotropy, does not change the degree of
polarization.

The CMB polarization is proportional to the dipole in the temperature
fluctuation at recombination, which in turn, being proportional to a time
derivative of the monopole, is proportional to $c_s$, in the region where
$\Delta_{T0}\propto\cos\phi$, with $\phi=k\int^\tau_0 d\tau c_s$. Since
$c_s\propto (1+R)^{-1/2}$ and $R\propto h^2 \Omega_b$, then the height of the
peaks in the polarization decreases for larger $\Omega_b$ (any other
dependence on $\Omega_b$ in eq. (\ref{soldp}) is not very significant).

More conclusions will be drawn in Sections IV and V, in the context of more
specific cosmological models. Now we turn our attention to the behaviour of
the temperature and polarization fluctuations at smaller scales, where the
tight coupling approximation starts to break down.

\subsection{Diffusion damping}

The approximations that lead to eq. (\ref{monopol2}) break down for scales much
smaller than $\tau_C$ ($k\tau_C>>1$), when the coupling between photons and
electrons is not so tight. To find the qualitative, and approximately
quantitative, behaviour at small scales it is necessary to expand the
temperature, polarization and velocity fluctuations to second order in
$\tau_C=\dot\kappa^{-1}$.  For these scales the role of gravity is
unimportant, so we solve equation (\ref{Boltzmann}) neglecting the
gravitational potentials\cite{Hu94,Peebles80}. Assuming solutions of the form
\begin{equation}
\Delta_T(\tau)=\Delta_T e^{i\omega \tau}\quad,\quad
\Delta_P(\tau)=\Delta_P e^{i\omega \tau}\quad,\quad
V_b(\tau)=V_b e^{i\omega \tau}\quad ,
\end{equation}
substituting this ansatz into the evolution equations (\ref{BoltzmannC}), and
expanding to second order in $\tau_C$, we obtain for $\omega=\omega_0+i\gamma$
\begin{equation}\begin{array}{lll}
& \omega_0 = {k\over \sqrt{3(1+R)}}\equiv k c_s\\ & \\ & \gamma=
{k^2 \tau_C \over 6(1+R)^2}[R^2 +{16\over 15}(1+R)]\\
\end{array}\end{equation}
where we have defined $c_s$, the photon-baryon sound speed.

The most important lesson to derive from this result is that at small scales,
those such that $k>>k_D$ defined below, the CMB temperature fluctuations and
polarization are damped by an exponential factor $e^{-\bar\gamma}$, with
\begin{equation}
\bar\gamma={k^2\over k_D^2}\equiv k^2\int_0^\tau {d\tau \over \dot\kappa} {1
\over 6(1+R)^2} [R^2 +{16\over 15}(1+R)]\quad ,
\label{Silkdamp}
\end{equation}
where we have taken into account the evolution of $R$ and $\dot\kappa$ with
$\tau$. This is just Silk damping due to photon
diffusion\cite{Silk68,Peebles80,Hu94}.

It is worth to stress at this point that the rigorous derivation of the
damping factor $\bar\gamma$ requires to take into account the polarization
dependence of Thomson scattering, as we did here and was first pointed out by
N. Kaiser\cite{Kaiser83}. If the evolution equations
for $\Delta_T$ are solved neglecting polarization (ignoring $\Delta_P$ in eqs.
(\ref{Boltzmann})), then a factor 4/5 would appear instead of the factor 16/15
in expression (\ref{Silkdamp}) for $\bar\gamma$\cite{Peebles80,Hu94}.   The
polarization dependence is important because after each scattering the
radiation is partially polarized. The polarization acts as a source of
anisotropy and  the anisotropy as a source of polarization. This coupling
between anisotropy and polarization makes the fluctuations decay more rapidly
than if there were no polarization dependence\cite{Kaiser83}. The difference
is not insignificant. For a CDM dominated universe such that $R\ll 1$, before
recombination ($x_e =1$) we get
\begin{equation} k_D^{-2}\approx 10^{7}
(1-Y_P /2)^{-1} (\Omega_b h)^{-1} (1+z)^{-5/2}h^{-2} {\rm Mpc}^2
\end{equation}
which differs by about $20\%$ from the result obtained
neglecting that the radiation is polarized\cite{Hu94}.

\section{Correlation Function and Total Polarization}

In this section we perform the multipole expansion of the CMB polarization
correlation function, and evaluate the total polarization produced by the
scalar fluctuations. In the previous section we evaluated the polarization
observed in a direction $\hat n$, $\Delta_P(\hat n,\vec k)$, induced by just
one single Fourier mode of the gravitational fluctuations. We were able to
choose Stokes parameters $Q\equiv\Delta_P$ and $U=0$ by the choice of the
$\hat z$ axis along the direction of $\vec k$, and defining the Stokes
parameters through projections of the CMB intensity along $\hat\theta$ and
$\hat\phi$.  But we can not do the same for all wavevectors. Now we must fix,
for each direction of observation, a pair of orthogonal axis, the same for all
wavevectors, to project the CMB intensity and evaluate the Stokes parameters.
After rotation of the $\vec k$-dependent basis to the fixed one, for each
direction of observation, there is a relatively simple relation between $Q$
and $U$ with $\Delta_P$ as calculated in the previous section\cite{Bond87}.
Let $\hat n$ be a direction of observation on the sky. The Stokes parameters
induced by a given Fourier mode of wavevector $\vec k$ read, in terms of
$\Delta_P$ as calculated in the previous section, as
\begin{equation}
Q(\hat n)=\Delta_P(\hat n,\vec k)\cos(2\phi_{\vec k})\quad ;
\quad U(\hat n)=\Delta_P(\hat n,\vec k) \sin(2\phi_{\vec
k})\quad ,
\label{qu}
\end{equation}
where $\phi_{\vec k}$ is, for a given direction of observation, the angle of
rotation between the two basis.

Given two directions of observation $\hat n_1$ and $\hat n_2$ the polarization
correlation function is defined as
\begin{equation} C_P(\hat n_1,\hat n_2)=\langle Q(\hat
n_1)Q(\hat n_2)+U(\hat n_1) U(\hat n_2)\rangle
\end{equation}
where $\langle\dots\rangle$
is an ensemble average. We assume that the density perturbations are gaussian.
Thus
\begin{equation}
\langle \Delta_P^*(\hat n_1,\vec k)\Delta_P(\hat n_2,{\vec
k}^{\prime}) \rangle = \vert \beta(k)\vert^2 (1-\mu_1^2) (1-\mu_2^2) e^{i k
r (\mu_1-\mu_2)}\delta^3(\vec{k}-{\vec k}^{\prime})
\end{equation}
where $\mu_1$ and $\mu_2$ are the cosines of the angles formed by $\vec k$
with $\hat n_1$ and $\hat n_2$ respectively,  $r\equiv\tau_0-\tau_D$ is the
distance to the last scattering surface, and we have written the result in
terms of the quantity $\beta(k)$ as defined in eq. (\ref{soldp}).

The correlation function can be written as
\begin{equation}
C_P(\hat n_1,\hat n_2)=
\int d^3 k \vert \beta(k)\vert^2 (1-\mu_1^2) (1-\mu_2^2)
e^{i r \vec{k}\cdot(\hat n_1-
\hat n_2)} \cos [2(\phi_{1\vec k}-\phi_{2\vec k})]
\label{correl}
\end{equation}
The multipole coefficients of the polarization correlation
function are given by
\begin{equation}
\langle a_l^2 \rangle \equiv \sum_m \int\int d\Omega_1 d\Omega_2
Y_{lm}^*(\hat n_1) Y_{lm}(\hat n_2)C_P(\hat n_1,\hat n_2)
\end{equation}

In the previous section we have seen that the most interesting structure in
the polarization of the CMB occurs for wavelengths comparable or smaller than
the horizon at decoupling. These wavelengths subtend an angle in the sky of
less than a few degrees. Small wavelengths have an effect on high multipoles
of the correlation function expansion, $l\gg 1$. Thus, we will make
simplifying approximations, valid for high multipoles only, the most
interesting at any rate. Recall that $\phi_{1\vec k}$ and $\phi_{2\vec k}$ are
the angles of rotation from the basis used to define the Stokes parameters
when we took the $\hat z$ axis in the direction of $\vec k$, to the  fixed
basis we use now. They are different for each direction of observation.
However, if $\hat n_1$ and $\hat n_2$ form a small angle $\theta$, then
$\cos[2(\phi_{1\vec k}-\phi_{2\vec k})]\sim 1 - {\rm O}(\theta^4)$. Thus,
replacing $\cos[2(\phi_{1\vec k}-\phi_{2\vec k})]$ by unity  in equation
(\ref{correl}) constitutes a very good approximation for high $l$. Under this
approximation  the multipoles have a simple expression
\begin{equation}
\langle a_l^2 \rangle\approx
{\pi\over 4} (2 l+1) \int d^3 k\vert \Delta_P(k)\vert^2 [c_{l+2}j_{l+2}(k r)+2
c_l j_l(k r) +c_{l-2}j_{l-2}(k r)]^2
\label{multip}
\end{equation}
where
\begin{equation}
c_{l+2}={4(l+1)(l+2)\over(2l+1)(2l+3)}\quad ,\quad
c_{l}={4 (l^2+l-1)\over(2l-1)(2l+3)}\quad ,\quad
c_{l-2}={4 (l-1)l\over(2l+1)(2l-1)}\quad .
\end{equation}
Notice that for
large $l$,  $c_{l+2}$, $c_{l-2}$ and $c_{l}$  all tend to unity.

Another important quantity is the power spectrum of the total
polarization, $W(k)$  defined as
\begin{equation}
C_P(\theta=0)=\langle Q^2 +U^2\rangle =\int_0^{\infty} {dk\over k} W(k)
\end{equation}
$W(k)$ measures the contribution of each wavevector $k$ to the total
squared polarization.
{}From equation (\ref{correl}) we  see that it is given by
\begin{equation}
W(k)=2\pi k^3 \vert \beta (k)\vert^2 \int_{-1}^{1} d\mu (1-\mu^2)^2
\label{poltot}
\end{equation}
The total degree of linear polarization is $P=(\int dk W(k)/k)^{1/2}$.

\section{Polarization in a COBE-normalized $\Omega_0=1$ CDM model}

In this section we specialize the evaluation of the polarization power
spectrum and its correlation function multipoles to  a specific cosmological
model, namely a spatially-flat, $\Omega_0=1$ cold dark matter cosmology with
scale invariant, adiabatic scalar fluctuations normalized to the large angle
anisotropy measured by COBE, under the assumption that the baryonic energy
density is much smaller than the radiation energy density ($R\ll 1$) at the
time of decoupling. We will also compare our analytic results with previous
numerical calculations.

The main ingredient in eq. (\ref{soldp}) necessary to evaluate $\Delta_P(\hat
n,\vec k) $ is the value of the dipole in the temperature anisotropy around
decoupling, which in turn can be derived from the monopole  through the
relation
\begin{equation}
i k \Delta_{T1}=-(\dot\Delta_{T0}+\dot\Phi)\quad .
\end{equation}
The monopole itself satisfies the equation
\begin{equation}
\ddot \Delta_{T0}+{\dot R\over 1+R}\dot\Delta_{T0}+
{k^2 \over 3(1+R)}\Delta_{T0}=-\ddot \Phi-{\dot R\over 1+R}\dot\Phi
-{k^2 \Psi \over 3}\quad .
\label{monop}
\end{equation}
In models with low baryon content, such that $R\ll 1$ at the time of
recombination, a WKB approximation to the exact solution of eq. (\ref{monop})
is a good approximation for most wavelengths of interest\cite{Hu94}.
If we define
\begin{equation}
\Delta_0\equiv \Delta_{T0} + \Phi
\end{equation}
we get for $\Delta_0$, in the WKB approximation,
\begin{equation}
\Delta_0=a \cos(\omega_0\tau) + b
\sin(\omega_0\tau) +{2 \omega_0} \int_0^{\tau} d\tau^{\prime}
\sin[\omega_0(\tau-\tau^{\prime})] \Phi(\tau^{\prime})\quad ,
\label{solution}
\end{equation}
with $\omega_0=kc_s\approx k/\sqrt 3$. We have neglected anisotropic stresses,
so that $\Phi=-\Psi$. The constants $a$ and $b$ are fixed by the initial
conditions. For adiabatic fluctuations $a={3\over 2} \Phi(0)$ and $b=0$. In
order to calculate the integral above we need to know the time dependence  of
the gravitational potential. Rather than resorting to a full numerical
approach, we will make some simplifying analytic approximations. During matter
domination the potential $\Phi$ remains constant on all scales, and the
integral in (\ref{solution}) is easily performed. For wavelengths which were
outside the horizon at the time of matter-radiation equality the behaviour of
the potential is also very simple: it remains constant during both radiation
and matter domination, but the value during matter domination is $9/10$ times
the value during during radiation domination. In this case:
\begin{equation}
\Delta_0(\tau)=- {3\over 10} \Phi(0) \cos(\omega_0 \tau) +{18\over 10}
\Phi(0)\quad {\rm if } \quad   k\ll k_{eq}
\end{equation}
For wavelengths which entered the horizon before matter-radiation equality,
the situation is different, since the potential $\Phi$ oscillates and decays.
For times well before matter-radiation equilibrium there is an analytic
solution for the potential:\cite{Mukhanov92}
\begin{equation}
\Phi= 3\Phi(0) {\sin(\omega_0 \tau)- \omega_0 \tau \cos(\omega_0 \tau) \over
(\omega_0 \tau)^3}
\label{potrad}
\end{equation}
If $k\gg k_{eq}$ the contribution of the integral is negligible, and then
\begin{equation}
\Delta_0 (\tau)=-{15\over 10} \Phi(0) \cos(\omega_0 \tau)\quad {\rm if}\quad \
k\gg k_{eq}
\end{equation}
With the solutions above for $\Delta_0(\tau)$ in the two different regimes, we
can evaluate the temperature dipole from $ i k \Delta_{T1}=\dot\Delta_0$ and
then use (\ref{soldp}) to find the polarization produced in this specific
model by one single Fourier mode:
\begin{equation}
\Delta_P (k)=\left\{\begin{array}{ll}
&{3\over 10} \Phi(0) (0.51 \omega_0 \Delta\tau_D) \sin(\omega_0\tau_D)
 e^{i k \mu(\tau_0 -\tau_D)}\quad {\rm if}\quad
k\ll k_{eq}\\
& \\
&{15\over 10} \Phi(0) (0.51 \omega_0 \Delta\tau_D) \sin(\omega_0\tau_D)
e^{i k \mu(\tau_0 -\tau_D)}\quad {\rm if}\quad  k\gg k_{eq}\\
\end{array}
\right.
\label{spol}
\end{equation}
For intermediate wavelengths the solution is not simple, so we find an
approximate solution using eq. (\ref{potrad}) during the radiation dominated
epoch, and matching the potential to a constant during matter domination.

We also have to take into account the damping of the temperature fluctuations
and polarization at small wavelengths due to photon diffusion (Silk damping)
and due to the finite width of the last scattering surface. We have already
calculated the effect of Silk damping at times around decoupling. The
exponential damping factor, $\bar\gamma=(k/k_D)^2$ is given by eq.
(\ref{Silkdamp}). Since $k_D$ is time dependent, and varies very rapidly
during recombination because of the fast change in the ionization fraction,
its effect on the damping of the polarization should be averaged over the
width of the last scattering surface. The average net damping factor
$\langle\gamma\rangle$ should be approximately given by
\begin{equation}
e^{-\langle\gamma\rangle}=\int_0^{\tau_0} d\tau {d\kappa\over d\tau}
e^{-\kappa} e^{(-{k \over k_D})^2}
\label{gammabar}
\end{equation}
The other source of damping comes from the finite width of the last scattering
surface. The oscillations of  the imaginary exponential in equation (\ref{dp})
produce a cancellation in the integral for wavelengths smaller than
$\Delta\tau_D$. To take this effect into account, the phase of the
perturbation should  be averaged over the width of the last scattering
surface, and so we should replace $e^{i k\mu \tau}$ by
\begin{equation}
\int_0^{\tau_0} d\tau {d\kappa\over d\tau} e^{-\kappa} e^{i k\mu \tau}
\end{equation}
This factor depends on $\mu$, the cosine of the angle between the wavevector
$\vec k$ and the direction of observation $\hat n$. Photons moving in a
direction perpendicular to $\vec k$ do not suffer the damping due to the
finite thickness of the last scattering surface. To simplify the calculations,
we average the damping factor over the angles, before performing the multipole
expansion. Our results for small wavelengths will be qualitative, and only
approximately correct quantitatively. We denote by $f(k\Delta\tau_D)$ the
averaged damping factor due to the finite width of the last scattering
surface. It is given by
\begin{equation}
f(k\Delta\tau_D) = 1/2 \int^1_{-1} d\mu
\int_0^{\tau_0} d\tau {d\kappa\over d\tau} e^{-\kappa} e^{i k\mu (\tau-\tau_D)}
\end{equation}
To summarize, our final expression for the polarization produced by one single
Fourier mode is given by the product of the factor $e^{-\langle\gamma\rangle}
f(k\Delta\tau)$ times expression (\ref{spol}). The damping factors and the
behaviour at intermediate wavelengths do not have simple analytic expressions,
so we handle them numerically.

Using the result for $\Delta_P$, we evaluate the polarization power spectrum
$W(k)$ as given by eq. (\ref{poltot}). Figure 1 shows the result, with the
gravitational potential normalized to adjust the COBE-DMR measurement of the
quadrupole in the temperature anisotropy, under the assumption of scale
invariance for the scalar fluctuations. To normalize we have taken on large
scales ${9\over 10}\Phi(0)=\Phi(\tau_D)=Ak^{-3}$ with $({1\over 3}A)^2=4 !
\langle a_{T2}^2\rangle /5(4\pi)^2$, and $\sqrt{\langle a_{T2}^2\rangle}
=2\times 10^{-5}$ the quadrupole in the temperature anisotropy measured by
COBE. Figure 1 displays the main features of the CMB polarization.  Very long
wavelengths contribute little to the final polarization because of the
relatively short width of the last scattering surface and the very tight
coupling between photons and electrons prior to decoupling. For intermediate
scales there are oscillations which follow the oscillations  of the
temperature anisotropy dipole, since those are the ones that by ``free
streaming''  during the decoupling transition  originate the quadrupole
anisotropy that gives rise to polarization. Finally, the contribution to the
polarization by smaller scales decays due to Silk damping and due to the
cancellations produced by the finite width of the last scattering surface.

We have also calculated the polarization correlation function multipoles
according to eq. (\ref{multip}). Since the largest contribution and the most
interesting structure in the polarization power spectrum occurs at
intermediate wavelengths, which correspond to $l\gg 1$, we can approximate
$c_{l+2}$, $c_{l}$ and $c_{l-2}$ by unity, their large $l$ limit. In that case
the combination of Bessel functions appearing in eq. (\ref{multip}) can be
approximated by $({2 l\over k r})^2 j_l(k r)$. On the other hand due to the
large value of $r$, this Bessel functions wildly oscillate, and it is a good
approximation to replace them by their approximate average
\begin{equation}
\langle j_l^2(x)\rangle \approx
\left\{ \begin{array}{ll} & [2x(x^2 - l^2)^{1/2}]^{-1}\quad {\rm if}\quad
x>l\\
&0 \quad {\rm if}\quad x<l \\
\end{array} \right.
\end{equation}
The result for the multipoles of the polarization correlation function is
plotted in Figure 2. The main features of this plot are easily understood from
the shape of the power spectrum in Figure 1.

The results in Figures 1 and 2 are  comparable to the numerical
results obtained by Bond and Efstathiou in reference \cite{Bond87}, more
specifically to their figures (4b) and (7a). The qualitative structure of
peaks is well reproduced. Our quantitative  results for the heights and
locations of the first peaks are relatively good, but accurate only up to
20 - 30 \%. For instance, the first four peaks in our Fig. 1 appear at
$k\approx 24,56,100,125\  {\rm Gpc}^{-1}$ respectively,
with heights $W(k)\approx 0.1,0.6,1.35,1.6\times 10^{-11}$,
while in Fig. (4b) of ref.
\cite{Bond87} they appear at $k\approx 16,50,80,100\  {\rm Gpc}^{-1}$
with heights (after
adjustment by a factor 15.4 to account for different conventions and
normalization\cite{footnote3}) $W(k)\approx 0.1,0.6,1.6,2.6\times 10^{-11}$.
The first four peaks in our Fig. 2 appear at
$l\approx  180,420,710,1000$
respectively, with $l^2C_l\approx 0.4,2.0,4.0,4.5\times 10^{-11}$,
while in Fig. (7a) of ref. \cite{Bond87} they appear at
$l\approx 130,320,560,800$ with heights (after
adjustment by a factor 7.7 to account for different normalization)
$l^2C_l\approx 0.3,1.5,4.9,6.2\times 10^{-11}$.

The quantitative discrepancies in the location of the peaks can be
partially attributed to the simplifying approximation  made in this section
that the photon-baryon sound speed was constant. This quantity varies
by around 15\% from the big-bang to the time of decoupling. Neglect of its
time-dependence induces
an error in our calculation of the phase of the oscillations and thus in the
location of the peaks. Besides, we cannot expect a good quantitative agreement
of the peak heights
on scales comparable to the horizon at the time of matter-radiation
equality due to our crude approximation for the potential at these
wavelenghths. We can not expect good quantitative agreement for small
wavelengths either given the rough approximations we
made in this section to average the damping effects. Our analytic approach is
not appropriate to obtain accurate quantitative predictions: due to our
simplifying approximations we can not expect better than 20-30 \% accuracy in
our estimates of the power spectrum. The most
interesting structure of peaks, however, is well reproduced by our analytic
results, and shows that our approach can be reliably used to analyse the
physical mechanisms that lead to polarization of the CMB, and their dependence
upon cosmological parameters.

\section{Polarization of the CMB in open universes}

We have seen in Section II that the polarization of the CMB is produced during
the process of decoupling of matter and radiation, and is proportional to the
width of the last scattering surface $\Delta\tau_D$, and to the value of the
dipole in the temperature anisotropy at the time of decoupling, which in turn
is determined by the value of the gravitational potential $\Phi(\tau_D)$. The
anisotropy in the CMB temperature also depends on these quantities, but
differently. For instance, it is very insensitive on large angular scales to
the values of $\tau_D$ and $\Delta\tau_D$. It is the aim of this section to
show that because of this different dependence, the ratio between anisotropy
and polarization at relatively low multipoles is very sensitive, in an open
universe, to the value of the matter density $\Omega_0$. Besides, while the
polarization of the CMB is produced during the decoupling transition, the
present temperature anisotropy receives contributions not only from the Sachs-
Wolfe effect, which is proportional to $\Phi(\tau_D)$, but also from the
integrated Sachs Wolfe effect (ISW), due to the time dependence of the
gravitational potentials. In an open universe, or in a flat universe with a
cosmological constant $\Lambda$, the potentials depend on time, and so the ISW
effect gives an additional contribution to the anisotropy. We will show that
this also makes the ratio between anisotropy and polarization dependent,
upon $\Omega_0$.

We shall evaluate the ratio  between the multipoles of the
temperature fluctuations and of the polarization correlation functions,
and investigate its dependence
upon $\Omega_0$ for $\Omega_0\le 1$ in two special cases:
when there is no cosmological constant, and when $\Omega_0 +
\Omega_{\Lambda}=1$,
with $\Omega_{\Lambda}\equiv\Lambda/3 H_0^2$. In both cases we shall assume a
scale invariant spectrum of adiabatic density fluctuations.

It has been shown\cite{Kamionkowsky94} that for multipoles such that $l_{curv}
<l<l_{D}$, where $l_{curv} =\pi\sqrt{(1-\Omega_0)/ \Omega_0}$ and
$l_D=r/\tau_D$, with $r$ the distance to the last scattering surface, the
coefficients of the multipole expansion of the temperature fluctuation
correlation function can be approximated by
\begin{equation}
C_l^T=A (1+{g(\Omega_0)\over l}) I_l^T
\end{equation}
where $A$ is the normalization of the scale-invariant energy-density power
spectrum of fluctuations, $\Phi(\tau_D)=A k^{-3}$, and $I_l^T
\equiv [9\pi l (l+1)]^{-1}$, both independent of $\Omega_0$. The dependence
upon $\Omega_0$ through the function
$g(\Omega_0)$ originates in the ISW, and is given by
\begin{equation}
g(\Omega_0)=36 \pi \int_{\tau_{LS}}^{\tau_0} \left ( {dF\over d\tau}
\right )^2 (\tau_0- \tau) d\tau
\end{equation}
where $F(\tau)$ gives the time dependence of the gravitational potential,
$\Phi(\tau)= \Phi(\tau_0) {F(\tau)\over F(\tau_0)}$. The evolution function
$F$ is given by $F(\tau)= D(\tau)/ a$\cite{Hu94} with
\begin{equation}
D=H \int {d{a/a_0}\over (H a/a_0)^3}
\end{equation}
The Hubble constant $H$ satisfies
\begin{equation}
H^2=({a_0\over a})^4 {a+a_{eq} \over a_0+a_{eq}}\Omega_0 H_0^2 -({a_0\over
a})^2 K + {\Lambda \over 3}
\end{equation}
with $K=-1$ in an open universe and $K=0$ in a spatially flat model.

Our result of previous sections for the coefficients of the multipole
expansion of the polarization correlation function is
\begin{equation}
C_l^P={\pi\over 4}\int d^3k \left| \beta(k)\right|^2[c_{l+2}j_{l+2}(kr)+2
c_{l}j_{l}(kr)+c_{l-2}j_{l-2}(kr)]^2
\end{equation}
This result, which was derived in a spatially-flat universe, is also valid for
multipoles $l\ge l_{curv}$. Indeed, for the wavelengths that contribute to
these multipoles, the radial functions appropriate for an open universe at the
time of decoupling are well approximated by the spherical Bessel functions of
the spatially flat case\cite{Kamionkowsky94,Hu94}. Besides, the evolution of
the gravitational potentials until the time of decoupling is basically the
same as in a flat universe provided that $\Omega_0 z_D>>1$.  We shall thus
work under this assumption, and consider mutipoles such that
$l\ge l_{curv}$ only. We shall also restrict our attention to multipoles $l<<
l_D$, which correspond to those before the first peak in Figure 2, just to be
able to work analytically. The wavevectors that significantly contribute to
these multipoles are such that $k\tau_D<<1$, and for them we can approximate
\begin{equation}
\left|\beta(k)\right|^2 =[0.17 \Phi(\tau_D)\omega_0 \Delta\tau_D
\sin(\omega_O \tau_D)]^2\approx [0.17 \Phi(\tau_D)\omega_0^2 \Delta\tau_D
\tau_D]^2
\end{equation}
and the multipoles can be rewritten as
\begin{equation}
C_l^P=A \left( {\Delta\tau_D \tau_D \over (1+R)r^2}\right)^2 I_l^P
\end{equation}
with
\begin{equation}
I_l^P={16\pi^2 \over 3}\int dx x [c_{l+2}j_{l+2}(x)+2 c_{l}j_{l}(x)+c_{l-
2}j_{l-2}(x)]^2
\end{equation}
Thus, the ratio between the temperature fluctuation and the polarization
correlation functions multipoles can be written as
\begin{equation}
{C_l^T\over C_l^P}=\left ({r^2 (1+R) \over \Delta\tau_D \tau_D}\right )^2
(1+{g(\Omega)\over l}) B_l
\label{ratio}
\end{equation}
with $B_l$ independent of $\Omega_0$ and $H_0$.

We want to find the explicit dependence of this ratio upon $\Omega_0$.
The distance to the last scattering surface, $r$,
is given by
\begin{equation}
r=\int^{a_0}_{a_0\over (1+z_D)} {da\over {\dot a} a}
\end{equation}
When there is no cosmological constant this reduces to
\begin{equation}
r={2\Omega_0 z_D+(2\Omega_0 -4)\sqrt{\Omega_0 z_D+1}-1 \over H_0 a_0
\Omega_0^2 (1+z_D)}
\end{equation}
When $\Omega_0
z_D >>1$ it can be approximated by $ r\approx 2/ \Omega_0 a_0 H_0$.

The dependence of $\tau_D$ and $\Delta\tau_D$ upon $\Omega_0$ can
be easily found exploiting the
fact that $z_{D}$ and $\Delta z_{D}$ are approximately independent of
$\Omega_0$ and $\Omega_b$\cite{Jones85}. Taking this into account, for
$\Omega_0 z_D>>1$ we get
\begin{equation}
\Delta\tau_D={\Delta z_D \over a_0 \Omega_0 H_0 (1+z_D)^{3/2}}\quad ;\quad
 \tau_D={2 \over a_0 \Omega_0^{1/2} H_0 (1+z_D)^{1/2}}
\end{equation}

Finally, the function $g(\Omega)$, which can be shown to be function of
the combination $(1-\Omega_0)/\Omega_0$ only, is well approximated by
\begin{equation}
g(\Omega_0)\approx 4.87 {(1-\Omega_0)\over \Omega_0}
\end{equation}

Using the above expressions we get for the ratio of multipoles (\ref{ratio})
\begin{equation}
{C_l^T\over C_l^P}=N_lG(\Omega_0,l)\equiv N_l[1+{4.87
(1-\Omega_0)/\Omega_0\over l}] \Omega_0^{-2}
\end{equation}
where $N_l$ is a normalization factor, defined so that $G(\Omega_0,l)$
measures the ratio between the temperature and polarization correlation
function multipoles normalized to the value of the same ratio when
$\Omega_0=1$.

We plot $G(\Omega_0,l)$ in Figure 3 for $l=30$. It is clear that the dependence
upon $\Omega_0$ is very significant. The ratio changes by a factor
of order $40$
for $\Omega_0\sim 0.2$ and  $250$ for $\Omega_0\sim 0.1$.

We now repeat the calculation for a model such that
$\Omega_0+\Omega_{\Lambda}=1$. $g(\Omega_0)$ can be approximated in this case
by $g(\Omega_0)=0.33 [(1-\Omega_0)/\Omega_0]^{2.23}$. There is no simple
analytic approximation for $r$ in this case, but in can be evaluated
numerically and, when $\Omega_0 z_D>>1$  the result is well fitted by
$r\approx 2/ H_0 a_0 \Omega_0^{0.39}$. Finally the ratio
of anisotropy to polarization multipoles  is given in this case by
\begin{equation}
{C_l^T\over C_l^P}=N_l G(\Omega_0,l)\equiv N_l[1+{0.33 [(1-\Omega_0 )/
\Omega_0]^{2.23}\over l}]\Omega_0^{0.22}
\end{equation}
which has a much weaker dependence upon $\Omega_0$ than in the open universe.
The result is plotted in Figure 4 for a fixed value of $\Omega_b$ and $l=30$.

\section{Conclusions}

We have performed an approximate analytic evaluation of the polarization
induced in the CMB on a wide range of angular scales by Thomson scattering
prior to decoupling in the presence of density perturbations with adiabatic
initial conditions, in a model with standard recombination.  Eq. (\ref{soldp})
gives the polarization induced by one single Fourier mode of the density
perturbations, down to scales such that $k\Delta\tau_D\approx1$. On smaller
scales the finite width of the last scattering surface and photon diffusion
damp the polarization of the CMB. The exponential damping factor due to photon
diffusion (Silk damping), the same for the anisotropy and the polarization,
is given by eq. (\ref{Silkdamp}). We stress the fact, already pointed out in
Ref. \cite{Kaiser83}, that the damping factor in the anisotropy that one would
derive neglecting the polarization dependence of Thomson scattering would be
slightly incorrect: instead of the factor $16/15$ in eq. (\ref{Silkdamp}) one
would get a factor $4/5$\cite{Peebles80,Hu94}. In conclusion, accurate
calculations of the anisotropy on scales where the width of the last
scattering surface is relevant should always include the polarization
dependence of Thomson scattering.

Eq. (\ref{soldp}) displays the fact that today's polarization of the CMB was
basically produced at the time when the tight coupling which kept the photon
distribution isotropic in the electrons' rest frame started to break down
during recombination. The degree of polarization is proportional to the
quadrupole in the temperature anisotropy around decoupling, which in turn is
mainly due to the ``free streaming" of the dipole during the last few
scatterings. This is manifested in eq. (\ref{soldp}) through the
proportionality
of the polarization upon the dipole in the temperature anisotropy at
decoupling and upon the width of the last scattering surface. The faster
decoupling occurs, smaller the degree of polarization induced in the CMB.

Figures 1 and 2 display our results for the CMB polarization power spectrum
and correlation function multipoles respectively, for a $\Omega_0=1$ CDM model
with adiabatic, scale-invariant scalar fluctuations normalized to the COBE-DMR
measurement of the large angle CMB temperature anisotropy. We have assumed a
standard recombination history and that the baryon density was much smaller
than the radiation density around decoupling. The agreement in the height and
position of the first peaks with respect to previous numerical
computations\cite{Bond87} is very good.

If the ionization history was as assumed here, with a tight coupling between
CMB photons and electrons until last scattering, and no later reionization,
then the relative locations of the peaks in the temperature anisotropy and
polarization correlation functions are in a simple and definite relation, as
we discussed in section (II.B). Thus, measurement of the relative locations
of the peaks in the anisotropy and polarization may give additional clues to
the ionization history. Let us also stress again that the relative heights
of  the peaks in the polarization are less dependent on parameters such as
$\Omega_b$ than those of the temperature fluctuations. Indeed, depending on
the value of $\Omega_b$ it is possible that the heights of even-numbered peaks
of the anisotropy are suppressed with respect to odd ones, due to a partial
cancellation between the adiabatic oscillations and the Sachs-Wolfe effect.
This does not happen to the degree of polarization, since it is not affected
by the photons' redshift. The absolute height of the peaks in the
polarization, on the other hand, is weakly dependent upon $\Omega_b$;
they decrease with increasing baryonic density.

At last, but not at least, we have shown that the ratio between anisotropy and
polarization multipoles is very sensitive to the value of $\Omega_0$ in an
open Universe, as evidenced in Figure 3. This strong dependence upon
$\Omega_0$ arises because anisotropy and polarization have a different scale-
dependence. Roughly, the polarization induced by a Fourier mode scales as
$k^2\Delta\tau_D\tau_D$ with respect to the anisotropy induced by the same
mode, on relatively large scales. In an open Universe with a scale invariant
spectrum of density fluctuations, and just for geometric reasons, a change in
the spatial curvature significantly changes the relation between the
wavenumber $k$ of the density fluctuations and the value of the multipole $l$
to which that wavenumber contributes the most, significantly changing the
value of the ratio between anisotropy and polarization multipoles. The ratio
of anisotropy and polarization multipoles also dependeds upon $\Omega_0$ in an
open Universe due to the time dependence of the gravitational potentials,
which affects anisotropy through the integrated Sachs-Wolfe effect, while it
does not affect the polarization.

\section*{Acknowledgements}

This work was partially supported by grants from Universidad de Buenos
Aires and Fundaci\'on Antorchas. D.H. is also supported by CONICET.

\section*{Figure Captions}

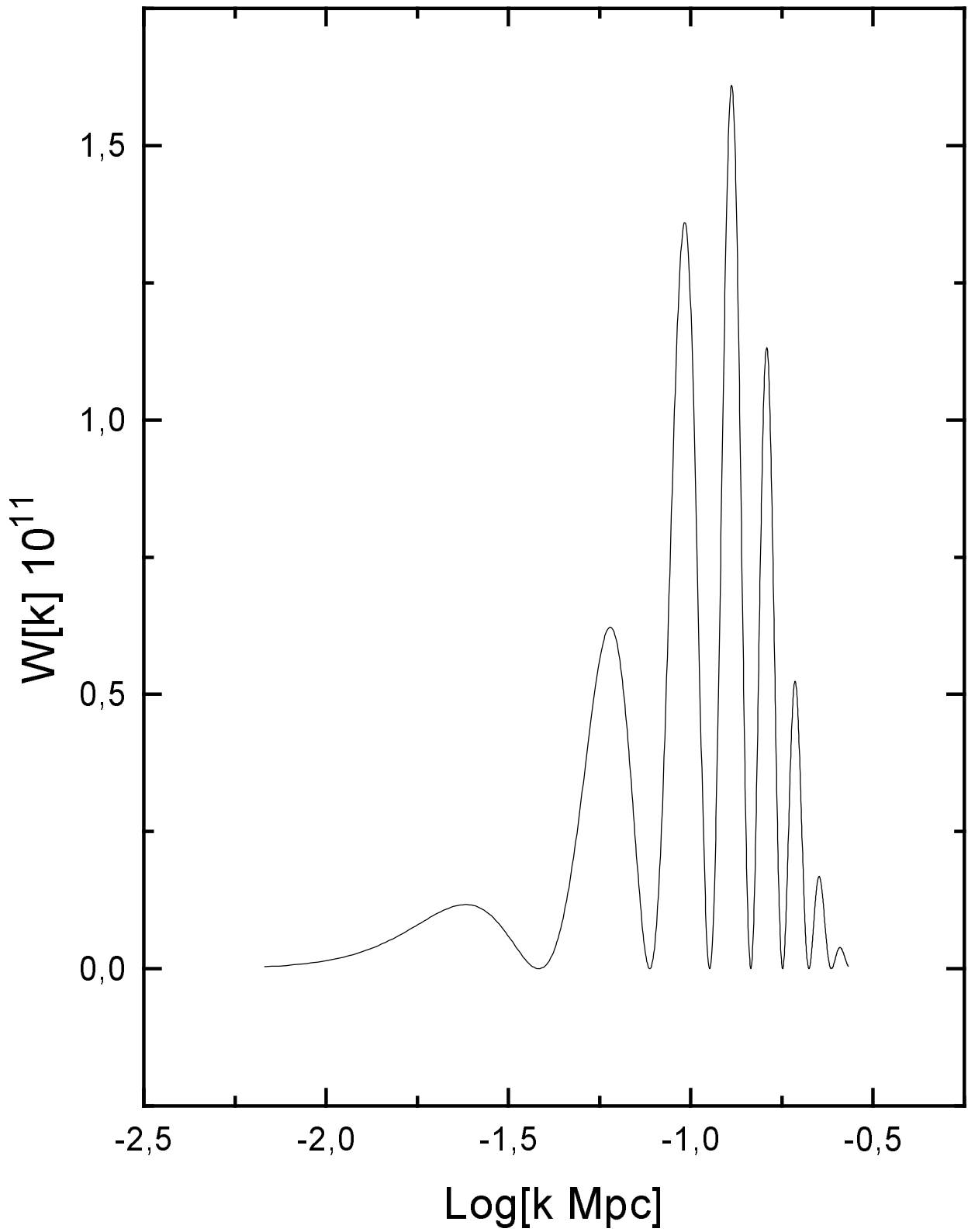
\noindent{\bf Figure 1:} Polarization power spectrum, as defined by eq.
(\ref{poltot}), normalized to the COBE-DMR measurement of the quadrupole
temperature anisotropy,  for a cold dark matter model with $\Omega_0 =1$,
$h=0.75$ and $\Omega_b=0.03$.

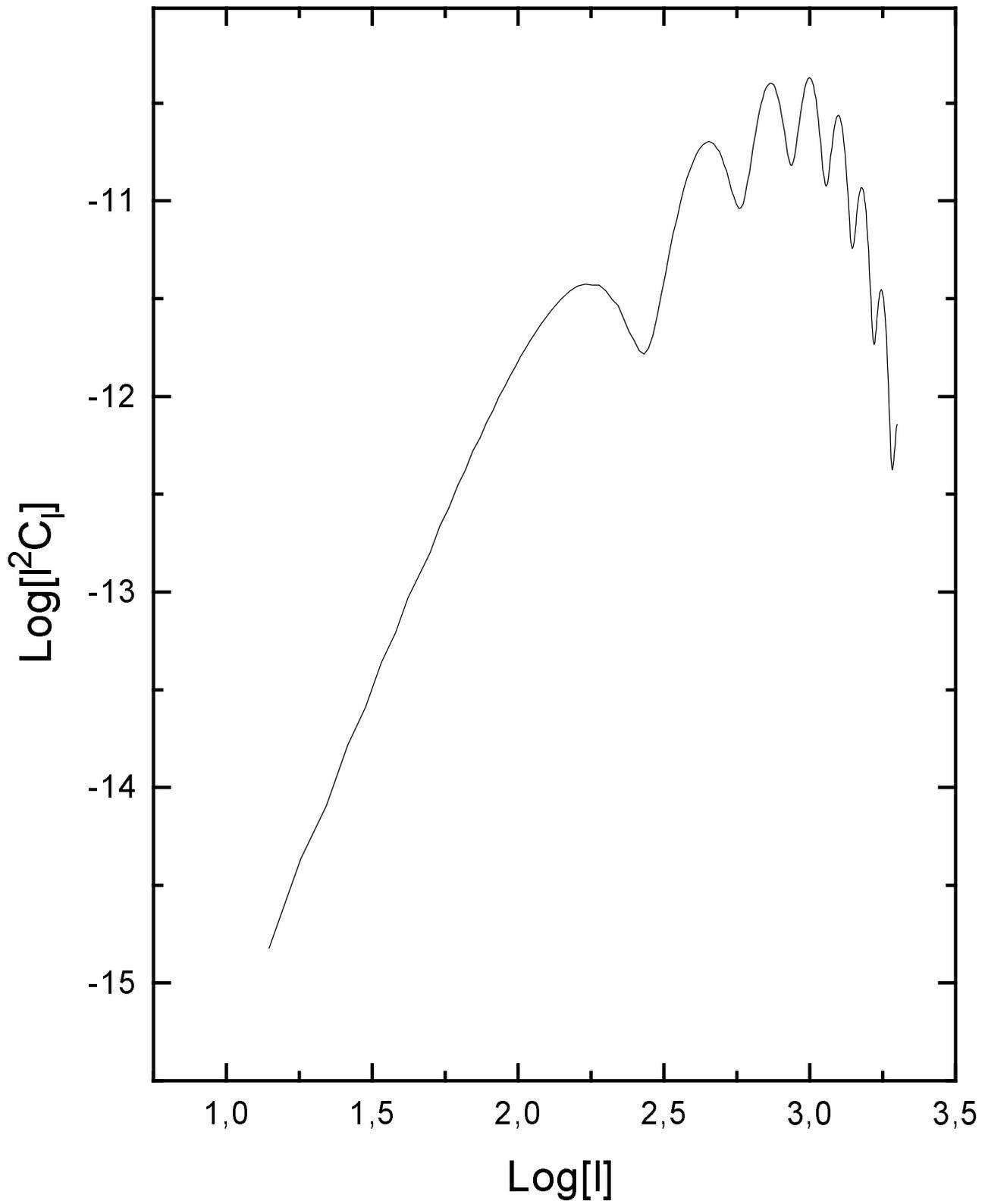
\noindent{\bf Figure 2:} Polarization correlation function multipoles
normalized to COBE for a cold dark matter model with $\Omega_0 =1$, $h=0.75$
and $\Omega_b=0.03$.

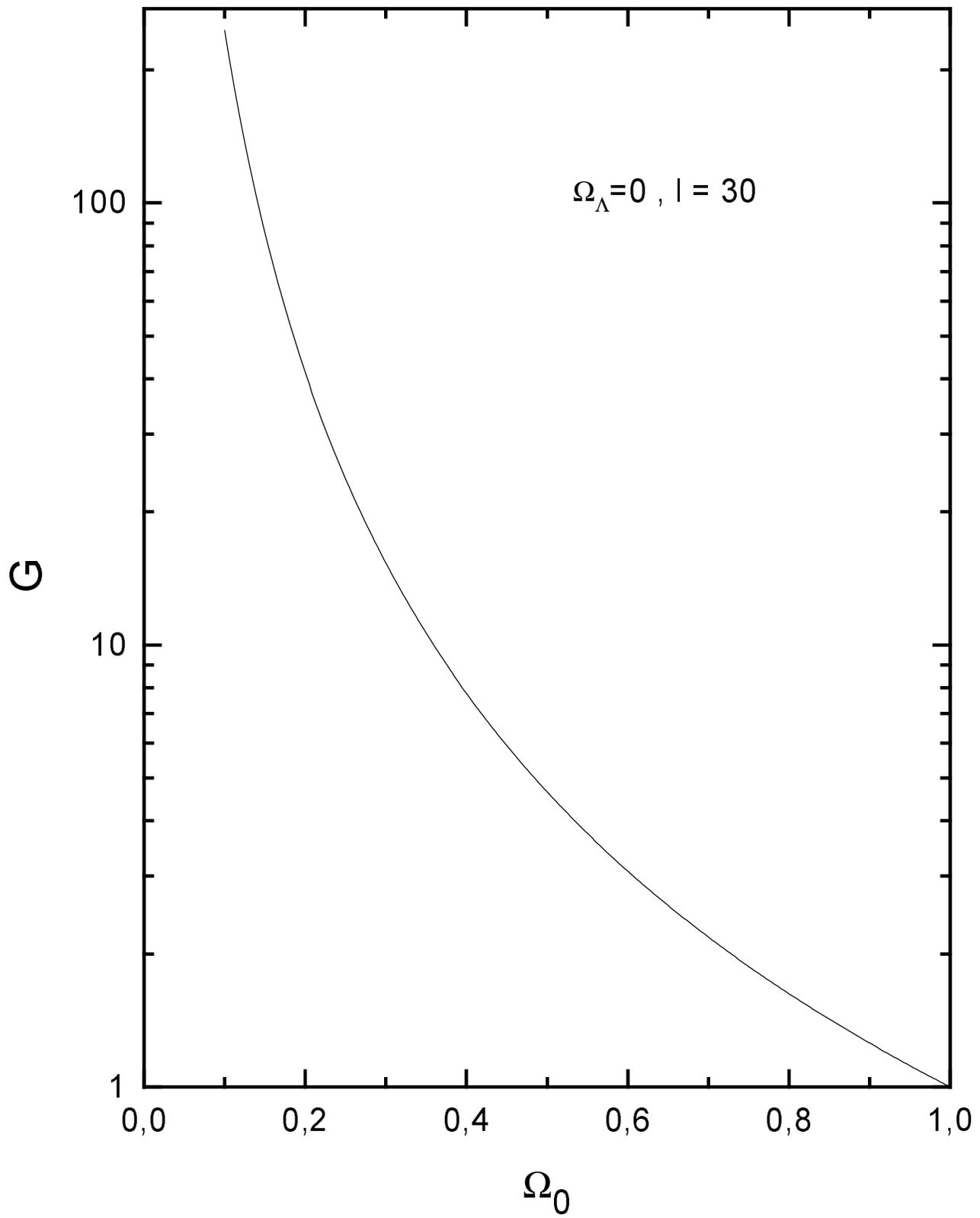
\noindent{\bf Figure 3:} Ratio of the anisotropy to the polarization
multipoles for $l=30$ as a function of $\Omega_0$ in
an open universe with no cosmological constant, normalized to its value when
$\Omega_0 =1$.

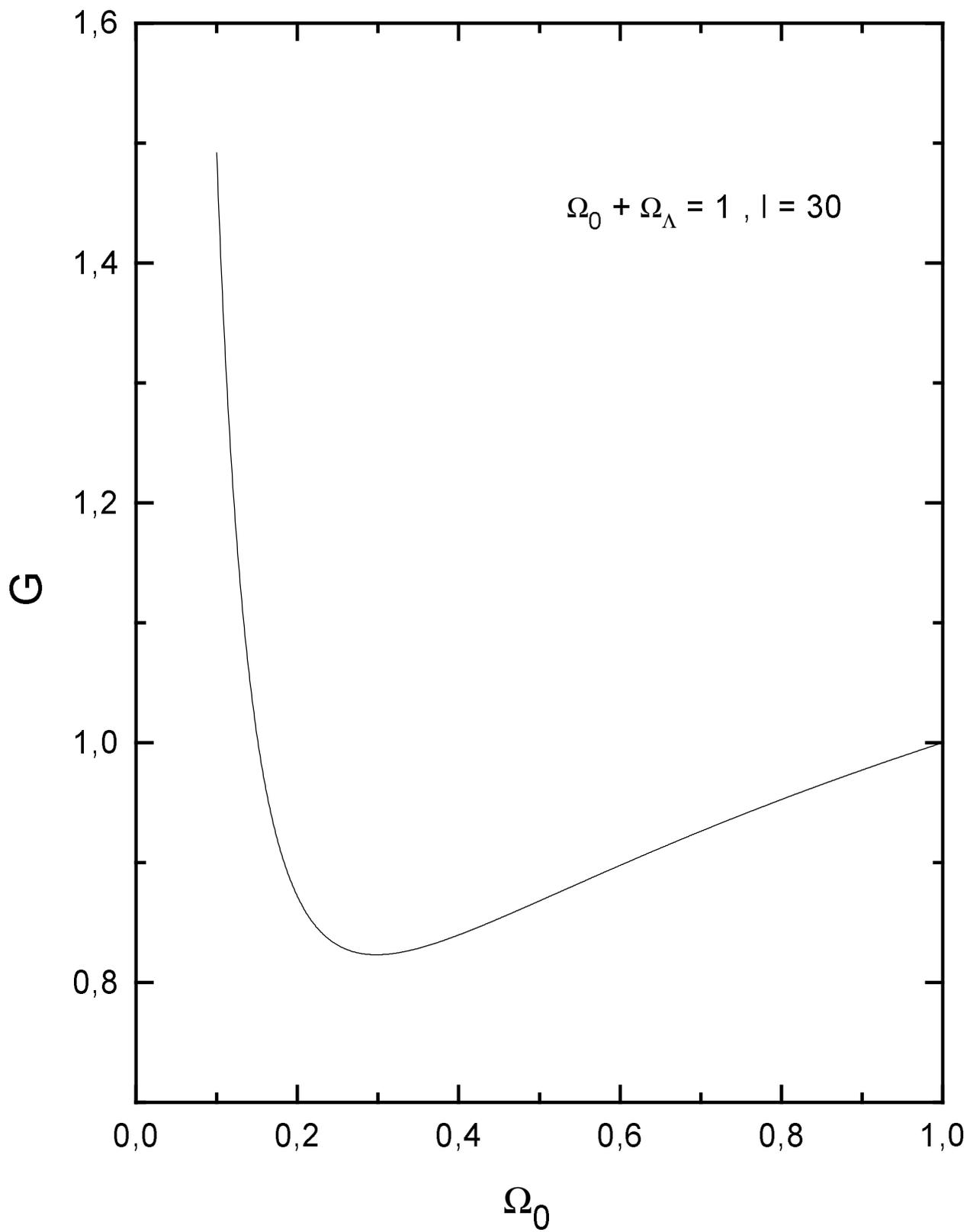
\noindent{\bf Figure 4:} Ratio of the anisotropy to the polarization multipoles
for
$l=30$ multipoles as a function of $\Omega_0$, normalized its value when
$\Omega_0 =1$, in a spatially-flat model with $\Omega_0 + \Omega_{\Lambda}=1$.


\begin{thebibliography}{99}
\bibitem{footnote1}Notice that our definition of $\Delta _T$ and $\Delta _P$
differs by a factor 4 from that of ref. \cite {Bond87}, while our definition
for the polarization correlation function $C_P$ turns out to be identical.
\bibitem{footnote2}Notice that our multipole expansion differs from that in
ref. \cite {Hu94} by a factor $(-i)^l/(2l+1)$.
\bibitem{footnote3}Notice that in order to compare our Figs. 1 and 2 with Figs.
(4b) and (7a) of ref. \cite{Bond87} some adjustments are needed to account for
different conventions and normalizations. Our W(k) corresponds to twice what
is plotted in (4b). Our definition of the quadrupole anisotropy $a_{T2}$ is 4
times that used in \cite{Bond87}. And finally the value of $a_{T2}$ used is
different.
\end{thebibliography}
\end{document}